\documentclass[lettersize,journal]{IEEEtran}
\IEEEoverridecommandlockouts

\usepackage{amsmath,amssymb,amsfonts}
\usepackage{graphicx} 
\usepackage{subfigure}
\usepackage{bm}
\usepackage{cite}
\usepackage{stfloats} 

\usepackage{booktabs}
\usepackage{tabularx}
\usepackage{threeparttable}

\usepackage{makecell}

\usepackage[ruled]{algorithm2e}
\usepackage{algorithmic}

\usepackage{bbding}

\usepackage{cuted}

\usepackage{color}

\usepackage{hyperref}

\ifCLASSINFOpdf

\else

\fi

\begin{document}

\title{Towards ISAC-Empowered Vehicular Networks: Framework, Advances, and Opportunities}

\author{Zhen~Du,~\IEEEmembership{Member, IEEE,}
Fan~Liu,~\IEEEmembership{Member, IEEE,} 
Yunxin~Li, 
Weijie~Yuan,~\IEEEmembership{Member, IEEE,} 
Yuanhao~Cui,~\IEEEmembership{Member, IEEE,}  
Zenghui~Zhang,~\IEEEmembership{Senior Member, IEEE,}  
Christos~Masouros,~\IEEEmembership{Senior Member, IEEE} and
Bo~Ai,~\IEEEmembership{Fellow, IEEE}


\thanks{Zhen Du is with the School	of Electronic and Information Engineering, Nanjing University of Information Science and Technology, Nanjing, 210044, China, and is also with the Department of Electronic and Electrical Engineering, Southern University of Science and Technology, Shenzhen 518055, China (email: duzhen@nuist.edu.cn).} 
\thanks{Fan Liu, Yunxin Li and Weijie Yuan are with the Department of Electronic and Electrical Engineering, Southern University of Science and Technology, Shenzhen 518055, China (email: liuf6@sustech.edu.cn; liyx2022@mail.sustech.edu.cn; yuanwj@sustech.edu.cn). (\textit{Corresponding author: Fan Liu})} 
\thanks{Yuanhao Cui is with the School of Information and Communication Engineering, Beijing University of Posts and Telecommunications, Beijing 100876, China (email: cuiyuanhao@bupt.edu.cn).}
\thanks{Zenghui Zhang is with the School of Electronic Information and Electrical Engineering, Shanghai Jiao Tong University,	Shanghai 200240, China (email: zenghui.zhang@sjtu.edu.cn).}
\thanks{Christos Masouros is with the Department of Electronic and Electrical
Engineering, University College London, London WC1E 7JE, U.K. (email: chris.masouros@ieee.org).}
\thanks{Bo Ai is with the State Key Lab of Rail Traffic Control and Safety, Beijing Jiaotong University, Beijing 100044, China. (email: boai@bjtu.edu.cn).}
}
\maketitle

\begin{abstract}
	Connected and autonomous vehicle (CAV) networks face several challenges, such as low throughput, high latency, and poor localization accuracy. These challenges severely impede the implementation of CAV networks for immersive metaverse applications and driving safety in future 6G wireless networks. To alleviate these issues, integrated sensing and communications (ISAC) is envisioned as a game-changing technology for future CAV networks.
	This article presents a comprehensive overview on the application of ISAC techniques in vehicle-to-infrastructure (V2I) networks. We cover the general system framework, representative advances, and a detailed case study on using the 5G New Radio (NR) waveform for sensing-assisted communications in V2I networks. Finally, we highlight open problems and opportunities in the field.
\end{abstract}

\begin{IEEEkeywords}
	integrated sensing and communications, vehicle-to-infrastructure, metaverse, beam tracking, 5G NR
\end{IEEEkeywords}

%
\IEEEpeerreviewmaketitle

\section{Introduction}
\subsection{Background and Motivations: ISAC-Empowered Vehicular Networks for Metaverse}
The metaverse has the potential to be a revolutionary application for future 6G mobile networks, bringing both significant technical advances and commercial opportunities. Connected and autonomous vehicles (CAVs) are mobile platforms that can benefit from the metaverse. The significance of the metaverse for CAV networks lies in providing a more intelligent, convenient, and safe travel experience, achieving seamless connectivity for CAV networks, improving transportation management efficiency, and bringing more intelligent and efficient development to future transportation and travel. To this end, critical Quality of Service (QoS) requirements are needed, including ultra-high throughput (at least 100 Gbps), high reliability (at least 99.9999\%), and ultra-low delay (less than 0.1 ms), etc.

\begin{figure*}[!t]
	\centering
	\includegraphics[width=7.1in]{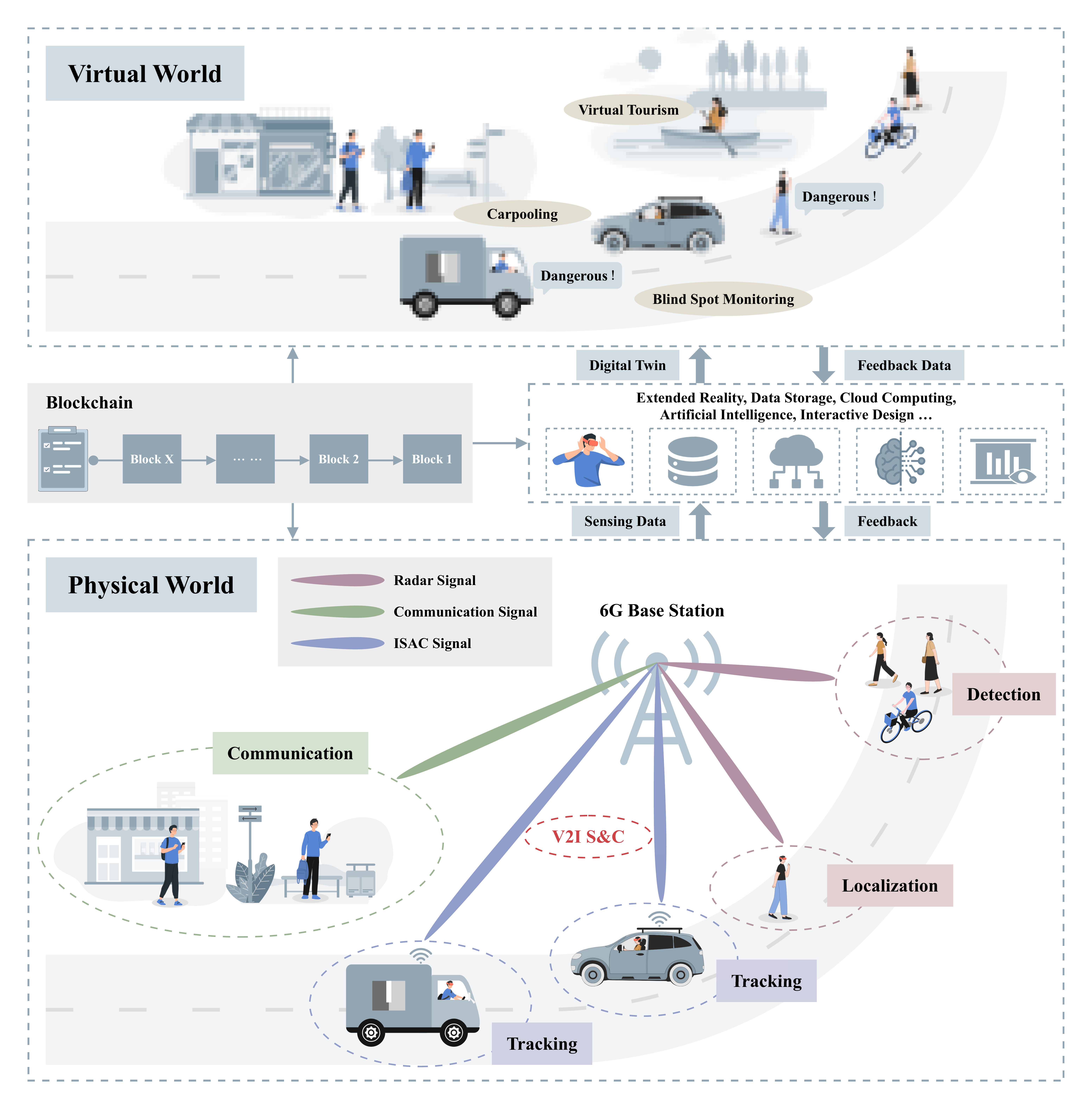}\\
	\caption{Illustration of metaverse of road environments in 6G mobile networks, where ISAC plays an important role.}\label{fig2}
\end{figure*}

As depicted in Fig. \ref{fig2}, the metaverse involves creating a virtual ecosystem that mirrors real-world road environments, utilizing a range of technologies such as extended realities, digital twins, artificial intelligence, large-scale distributed computing, interactive design and development tools, and more, to support its infrastructures \cite{tang2022roadmap}. 
To enable a seamless integration between the physical and digital worlds in the future 6G era, the metaverse requires ubiquitous and highly accurate wireless sensing services. 
One example of CAV networks is their ability to connect commercial vehicles in a shared platform within the metaverse, enabling carpooling and rental services. To achieve this, it is necessary to make unified driving decisions within the metaverse while ensuring the crucial sensing resolution, accuracy, and communication QoS related to the physical layer (PHY) performance in the real world.
This will unlock the potential for co-designing sensing and communications (S\&C) functionalities on a single platform, using the same spectrum and even a unified transmitting waveform. This approach, known as integrated sensing and communications (ISAC), will facilitate the mapping of virtual and real objectives \cite{cui2021integrating}, ensuring an immersive metaverse experience.
In an ISAC system, S\&C functionalities are no longer treated as separate end-goals, rather they are co-designed to pursue mutual benefits. The motivations of ISAC are influenced by both internal technological trends and external commercial requirements of S\&C, as summarized below.
	\begin{itemize}
		\item \textbf{Technical Trends:} Both S\&C are evolving towards higher frequencies, such as millimeter-wave (mmWave) and terahertz (THz) bands, where ISAC is a cost-effective paradigm shift which promotes the miniaturization of equipment, the improvement of spectral and power efficiencies, and the reduction of hardware and signal processing overheads, etc. 
		\item \textbf{Commercial Requirements:} The convergence of S\&C technologies has the potential to enhance spectral efficiency, resulting in significant economic benefits. Nevertheless, current 5G mobile networks are not equipped with basic sensing capabilities. Therefore, it is crucial for the next generation of mobile networks, 6G, to incorporate sensing abilities such as tracking, localization, mapping, and imaging.
	\end{itemize}
Based on motivations above, two primary advantages of ISAC can be summarized \cite{cui2021integrating}. Firstly, ISAC provides \textit{integration gains} by allowing shared radio resources for dual-functional S\&C, which helps reduce the duplication of transmissions, devices, and infrastructure. Secondly, ISAC offers \textit{coordination gains} by enabling mutual assistance between S\&C.

\begin{figure*}[!t]
	\centering
	\includegraphics[width=7.0in]{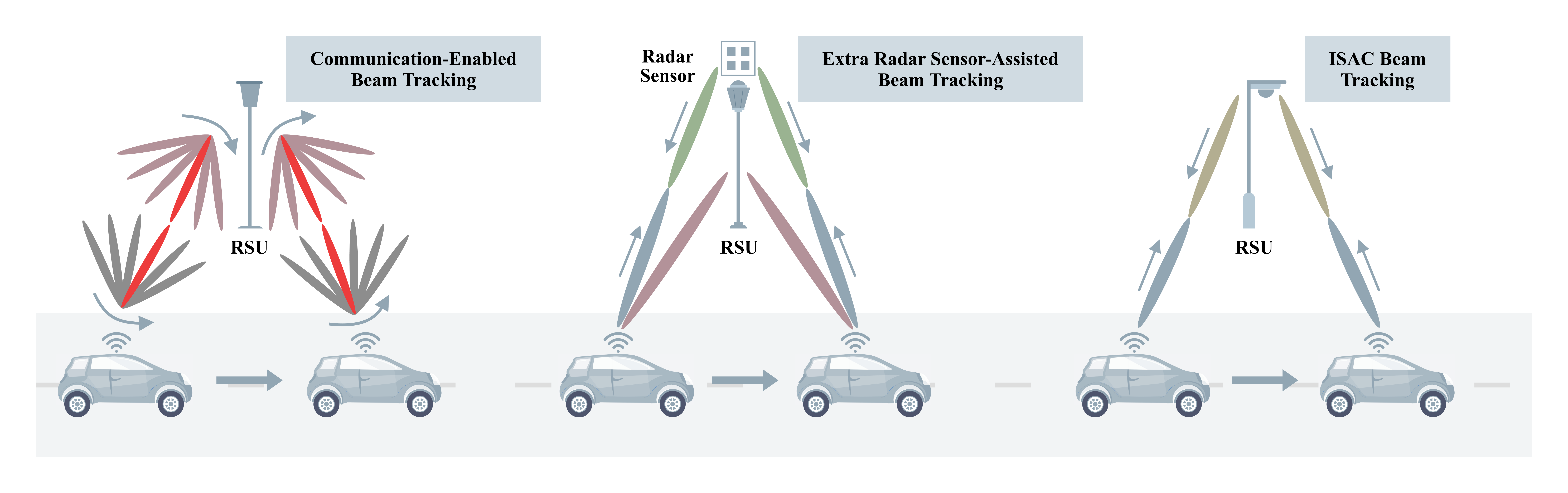}\\
	\caption{Illustration of three representative beam management frameworks.}\label{fig1}
\end{figure*}

\subsection{Research Progress: Classifications of Beam Management}

	The vehicle-to-everything (V2X) networks enable information sharing among vehicles, infrastructures, Internet, and pedestrians, among which the vehicle-to-infrastructure (V2I) links are expected to be tightly connected with 6G mobile networks. By exploiting massive multi-input-multi-output (mMIMO) antennas and increasing the carrier frequency to mmWave and THz bands, the communication throughput is capable of being immensely improved due to the large bandwidth provided by substantial spectral resources, and the significant beamforming gain of pencil-sharp beampattern with low sidelobes. To that end, the efficient mMIMO beam management becomes a basic premise and exhibits a more serious bottleneck challenge in high-mobility V2I networks due to fast-changing channel characteristics, which is regarded as a limiting factor for practical implementation of CAV metaverse applications. On the whole, representative beam managements include the following three frameworks, as shown in Fig. \ref{fig1}.
	
\begin{figure}[!t]
		\centering
		\includegraphics[width=3.5in]{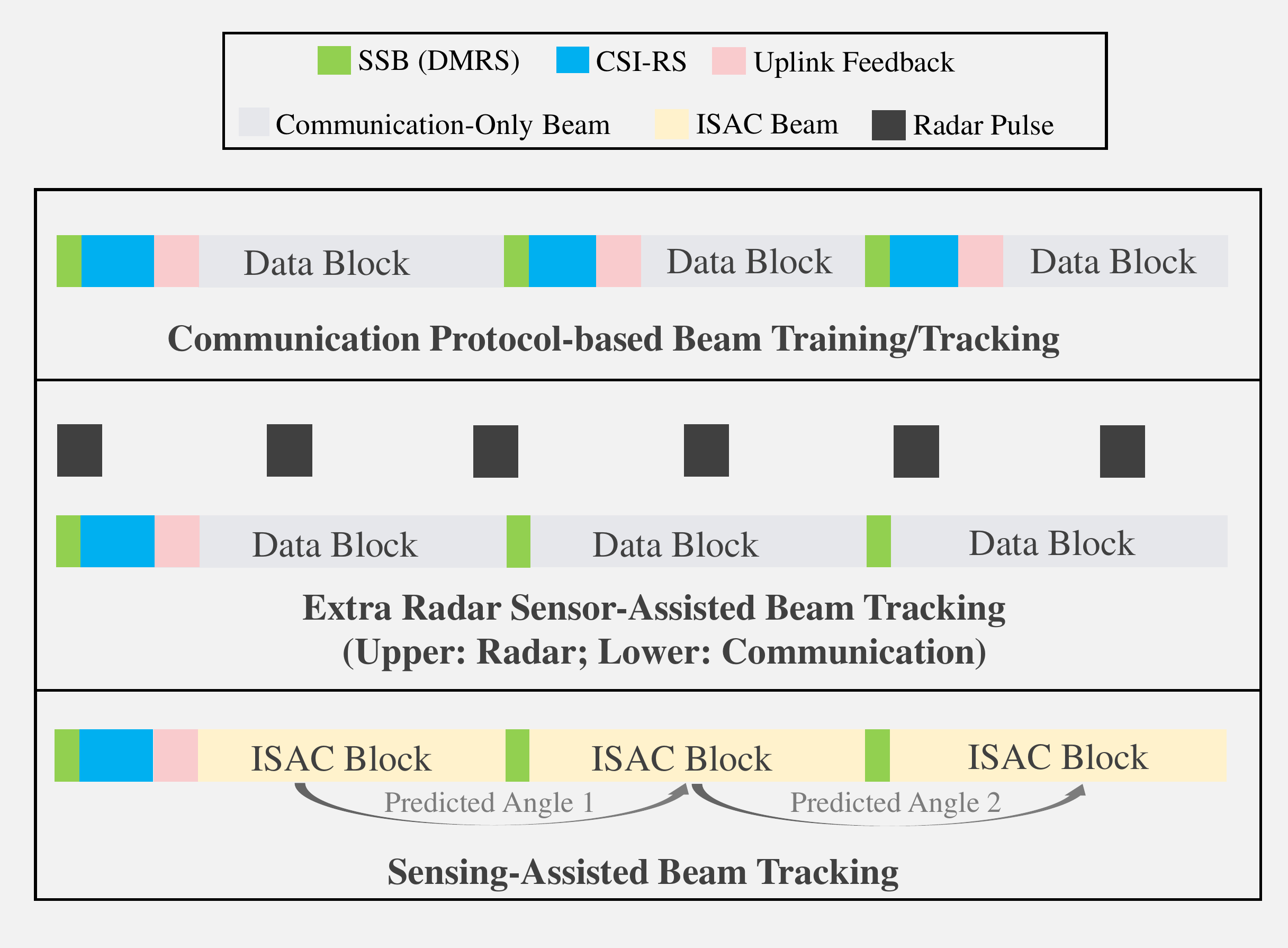}\\
		\caption{Illustration of frame structures: communication protocol-based beam training/tracking, extra radar sensor-assisted beam tracking, and sensing-assisted beam tracking.}\label{fig3}
	\end{figure}

	\begin{itemize}
		\item \textbf{Communication-Enabled Method:} In conventional V2I networks, beam management is achieved through beam training, which requires frequent signaling feedback between vehicles and base stations, and incurs significant angle searching overheads throughout the spatial domain. To this end, more pilots, such as synchronization signal blocks (SSBs) and channel-state-information reference signals (CSI-RSs), have to be allocated in the preamble of frames, which significantly reduces the spectral efficiency. While the beam tracking technique is capable of reducing the angular searching region through leveraging the temporal correlation between adjacent transmission slots, it still needs repeated handover and feedback to ensure seamless vehicular connectivity. As a consequence, current beam management techniques based on communication-only protocols are unable to meet QoS requirements for both CAV and the metaverse simultaneously.
	
		\item \textbf{Extra Radar Sensor-Assisted Method:} ISAC's integration and coordination gains enable independent S\&C components to be co-designed collaboratively. In the context of V2I networks, sensing-assisted communication can reduce frequent sweeping overheads, while improving localization accuracy and robustness in high-mobility scenarios. An initial concept for implementing sensing-assisted communications involves installing a collocated radar sensor to the roadside unit (RSU) \cite{2016Millimeter}. This sensor would be used to gather channel state information, which can be utilized to aid for beam alignment in V2I networks. While reducing training overheads may improve the S\&C coordination gain, incorporating an additional radar sensor with its own spectral band, waveform generator, and processing unit may lead to a loss of integration gain. This may increase spectral consumption, hardware cost, and equipment size, ultimately resulting in an approach that may not be cost-effective.

		\item \textbf{ISAC Method:} One approach that exhibits more promise is to use the communication signal itself for sensing, rather than employing only the preamble. This can be achieved by utilizing the entire data frame. Essentially, the sensing receiver of the RSU can extract localization information from echoes using radar sensing algorithms. Compared to communication-only approaches, the ISAC-enabled sensing-assisted communications approach offers several advantages \cite{liu2020TWC,yuan2020,yuan2021integrated,du2021integrated,meng2022vehicular}, including:

		\begin{itemize}
		\item[1] \textbf{Release of Pilot Overheads:} 
		The use of an ISAC-enabled approach can alleviate the burden on CSI-RS and uplink feedbacks, as localization information can be obtained through echoes at the sensing receiver. It is important to note, however, that demodulation reference signals (DMRS) are still necessary for coherent demodulation in the communication receiver and are therefore included as part of the SSB. To further illustrate this point, a comparison of frame structures can be found in Fig. \ref{fig3}.
		\item[2] \textbf{Significant Localization Gain:} 
		Although sensing echoes experience larger path loss due to the round trip as compared to the communication propagation in a single trip, the benefits of using a longer block for localization are still significant. Using an entire block instead of limited pilots results in a remarkable improvement in the signal-to-noise ratio (SNR) due to matched filtering, which is less affected by path loss. Moreover, the additional degrees-of-freedom (DoFs) of distance and velocity, beyond just angle, also contribute to the SNR gain. As a result of these factors, the overall localization accuracy is significantly improved.
		\end{itemize}	

	\end{itemize}

	The remainder of this article is organized as follows.
	Firstly, we start by holistically reviewing the systematic framework of ISAC-enabled V2I networks. Subsequently, we introduce the main challenges and their solutions. Furthermore, a case study is elaborated by designing ISAC frame structures and sensing-assisted beam management methods based on 5G NR protocols. Finally, we shed light on a number of open problems and opportunities deserved to be further studied in the future.

\section{General System Framework}
	\subsection{{Fundamental Architecture}}
	We consider mmWave RSU with active sensing capabilities, specifically focusing on the collocated ISAC transmitter and sensing receiver. The ISAC transmitter is particularly important because it concurrently transmits a unified waveform for both S\&C functionalities. To ensure effective operation of the system framework, it is crucial to maintain good isolation between the ISAC transmitter and sensing receiver, especially since the RSU transmits a continuous signal. This necessitates that the RSU operates in in-band full-duplex (IBFD) mode to eliminate self-interference. Current techniques are capable of achieving self-interference cancellation of more than 110 dB, which is widely acknowledged \cite{barneto2019full}. Additionally, it is noted that, the RSU utilizes a mMIMO uniform linear array (ULA) or uniform plane array (UPA) to generate a pencil-sharp narrow beam with exceptionally high array gain.

	\subsection{{ISAC Transmit Waveforms}}
		\begin{itemize}
			\item \textbf{Orthogonal Frequency Division Multiplexing (OFDM):} OFDM modulation has been widely applied in 4G/5G cellular systems and WiFi networks, thanks to its numerous benefits such as resilience to multipath interference, simple structure and low implementation cost facilitated by the fast Fourier transform (FFT) algorithm, orthogonality of subcarriers, and frequency diversity, etc. Moreover, OFDM has gained much attention in the radar community due to its unique properties, such as a thumbtack ambiguity function and the ability to synthesize a large time-bandwidth product. However, it is susceptible to severe Doppler spread. This becomes particularly relevant in CAV networks where high-mobility is a key factor. The sensitivity of OFDM to Doppler effects, including intercarrier interference (ICI), may lead to beam tracking failures. 
			\item \textbf{Orthogonal Time-Frequency Space (OTFS):} OTFS is a modulation scheme that operates in the two-dimensional domain of delay-Doppler (DD), enabling it to adapt to the high dynamics of time-varying channels in the time-frequency (TF) domain. The ingenious design of OTFS enables a perfect interaction between transmitted signals and the DD domain channel characteristics. By converting information modulation to the DD domain, OTFS can effectively mitigate the effects of channel fading and interference. This is also crucial for exploiting echoes to estimate delay and Doppler shifts. Compared to OFDM, OTFS offers full diversity as each data symbol is spread over the entire TF domain and has a lower peak-to-average-power ratio (PAPR), making it a better candidate for certain applications \cite{yuan2021integrated}.
		\end{itemize}	
	
	\subsection{{S\&C Channels at MmWave Bands}}
	Overall, both S\&C exhibit similar channel characteristics as they operate in common mmWave bands. Firstly, the mmWave channels for both S\&C are sparse, meaning that they involve fewer non-line-of-sight (NLoS) components compared to conventional sub-6GHz channels. Secondly, the lower sidelobes of mmWave beams for both S\&C contribute to a reduced vulnerability to eavesdropping and jamming.
	Moreover, the ISAC transmitter sends a unified signal to the vehicle, 
	where the impulse response of sensing channel is determined by the target being sensed. 
	Many studies treat the vehicle as a point-like object, which is usually based on the slowly fluctuating Swerling I-type model. However, this model of the sensing channel is inadequate when the resolution is higher. For example, when using the mMIMO array in mmWave bands, the sensed target has strongly extended contents.
	Despite mmWave S\&C channels having higher power attenuation than sub-6GHz channels, the high array gain of mMIMO beams can effectively counteract this path loss. Additionally, NLoS paths contribute to the communication throughput, while NLoS clutter components received by the sensing receiver may jeopardize the sensing capability. The tradeoff between these factors will be further examined in the following.

	\subsection{{Sensing Receive Algorithms}}
		\begin{itemize}
			\item \textbf{Kalman Filtering:} The Kalman filter, which is a linear minimum mean square error (MMSE) estimator, uses a series of measurements observed over time, including statistical noise and other uncertainties, to estimate unknown variables. It does so by estimating a joint probability distribution of the measurement and state variables for each time slot. Additionally, it determines the optimal weight factor between the measurement model and state evolution model, known as the Kalman Gain. For non-linear models, the linear approximation can be utilized, such as with extended Kalman filtering (EKF) and unscented Kalman filtering (UKF), among others. The Kalman filter family is especially useful for high-mobility V2I communications due to its predictive beamforming capabilities \cite{liu2020TWC}. Based on its advantages, including its ability to handle non-stationary data, the Kalman filter family is a promising technique for beam management applications. However, it is important to note that applying Kalman filtering may result in non-optimal performance in the presence of non-Gaussian noise. Additionally, successful implementation of Kalman filtering requires prior knowledge of measurement and state errors, which can be estimated beforehand or substituted with the corresponding Cr$\acute{\text{a}}$mer-Rao bound (CRB) values.
			\item \textbf{Bayesian Filtering:} Kalman filtering is a linear filtering technique that is considered optimal for estimating states in systems with Gaussian uncertainties. It is a special case of Bayesian filtering, which involves estimating the probability distribution of unknown variables based on sensor measurements and prior knowledge. From an optimal estimation perspective, sensing-assisted beam tracking involves inferring unknown variables such as distance, velocity, and angle from measurements and states using the maximum \textit{a posteriori} (MAP) estimator. The motion parameters of vehicles can be estimated and predicted using a factor graph and the message passing algorithm. While the standard message passing algorithm cannot provide closed-form solutions, a modified mean field message passing algorithm and second-order Taylor expansion can be used to linearize the inverse trigonometry functions \cite{yuan2020}. This approach is preferable to particle filtering, which requires a large number of particles and can be computationally expensive. It is also worth highlighting that using the mean field algorithm and Taylor expansion balances complexity and performance effectively.
		\end{itemize}


\section{Challenges and Solutions}
The constructed ISAC-based system framework has displayed significant potentials in enabling sensing-assisted communications for V2I networks. However, some key obstructions still prevent its practical implementation. 
To summarize, we list some critical issues that we have solved as below.

	\subsection{{ISAC Resource Allocation}}
	At the macro level, one of the primary issues faced by ISAC-enabled V2I networks is resource allocation. The allocation of transmit resources to multiple services is crucial for achieving better QoS of S\&C. However, due to the limited resource budget, such as power and bandwidth, there is a tradeoff in performance among S\&C users. To address this issue, we emphasize the importance of allocating bandwidth to different beams, as this approach can help to mitigate mutual interference between adjacent beams, particularly when they are partially overlapped.
	To illustrate this, a unified resource allocation framework has been proposed \cite{dong2022sensing}, where the QoS of sensing, including detection, localization, and tracking, is quantified by specific metrics. This framework aims to optimize the allocation of resources by prioritizing the most critical metrics for each type of service, thereby improving the overall performance of ISAC-based vehicular networks.
	
	\begin{itemize}
		\item \textbf{Probability of Detection:} When a threshold is set for a specific probability of false alarm, the Neyman-Pearson criterion can be used to measure how detectable a target is.
		\item \textbf{CRB:} The CRB is known as a theoretical lower bound on the variance of an unbiased estimator. Usually the mean square error (MSE) is able to approach the CRB in the high-SNR regime.
		\item \textbf{Posterior CRB (PCRB):} The PCRB is naturally tailored for tracking services since the filtering procedure is contributed by both the measurement and the prior state evolution model.
	\end{itemize}

	By combining the three sensing QoS metrics with communication QoS, such as achievable sum-rate, it is possible to achieve adjustable performance trade-offs for resource allocation by allocating power and bandwidth resources according to the desired system requirements. Fortunately, the resource allocation problems that arise from this framework can be effectively solved by alternately optimizing the power and bandwidth allocation variables, thanks to the convexity of the formulated problems \cite{dong2022sensing}.

	\subsection{{Point Targets versus Extended Targets}}\label{subsec3B}
\begin{figure}[!t]
		\centering
		\includegraphics[width=3.5in]{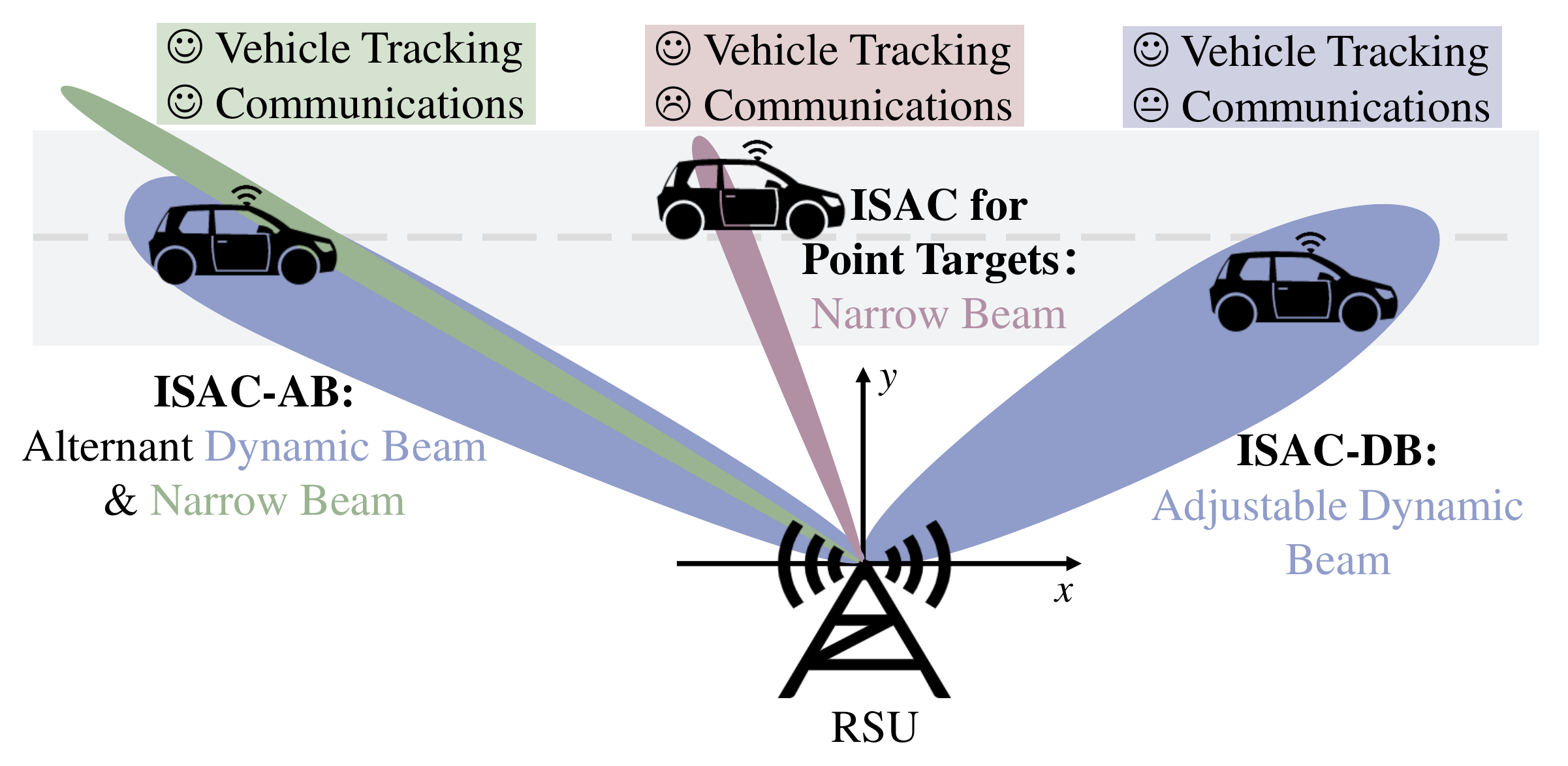}\\
		\caption{Illustration of beam tracking for extended targets and point targets.}\label{fig7}
	\end{figure}

\begin{figure}[!t]
		\centering
		\includegraphics[width=3.35in]{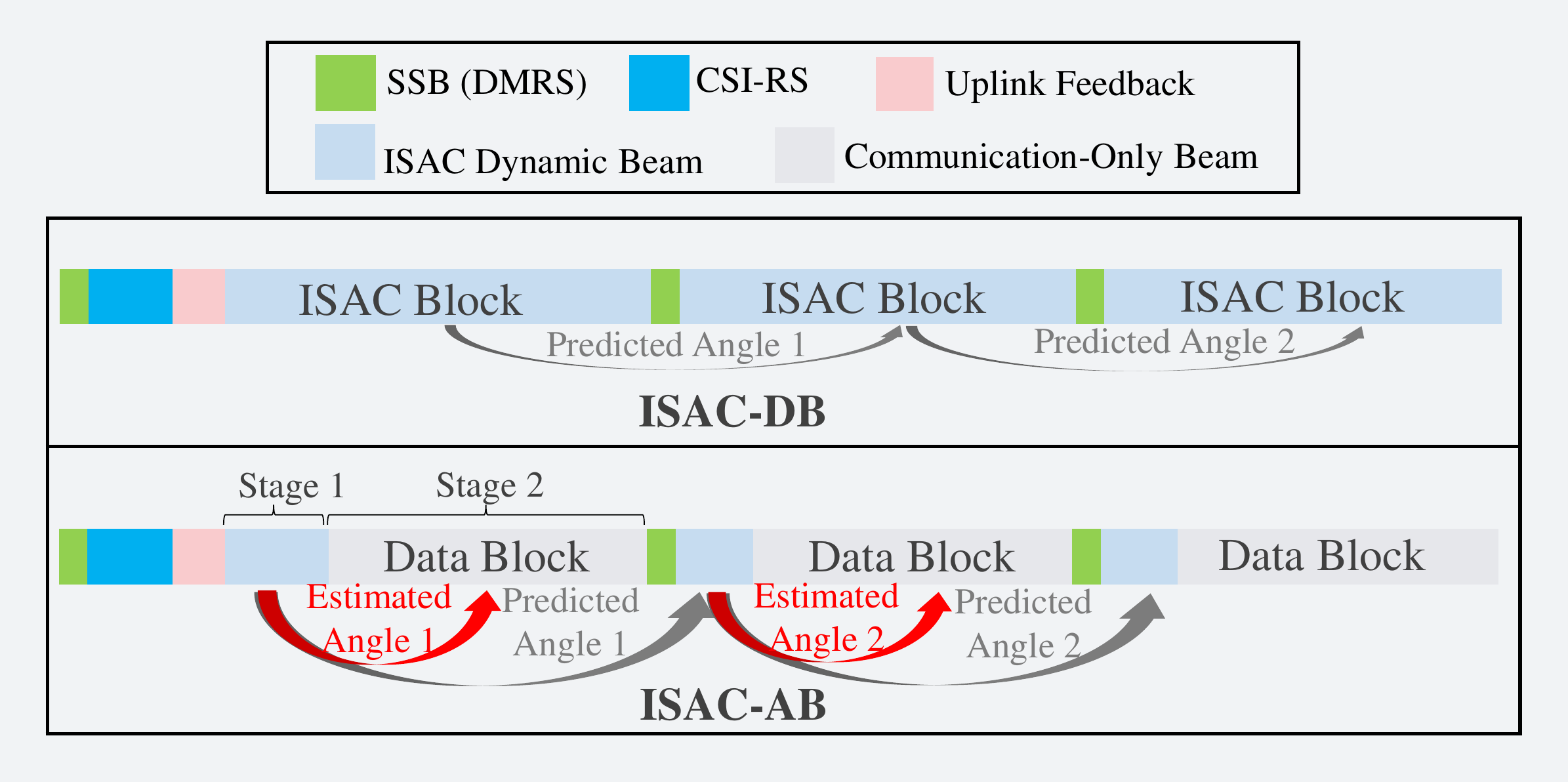}\\
		\caption{Tailored frame structures of ISAC-DB and ISAC-AB.}\label{fig4}
	\end{figure}

	Thanks to the abundant spectral resources available at mmWave bands for radio transmission, the sensing range resolution is relatively high. Moreover, the pencil-sharp mMIMO beam may not illuminate the entire vehicle, resulting in highly extended contents at the sensing receiver in both angular and range domains. This phenomenon leads to an undesirable outcome where the RSU cannot identify the precise location of the communication antenna, even when the vehicle is accurately tracked, as illustrated in Fig. \ref{fig4}. Furthermore, the communication antenna may actually be outside the effective mainlobe of the beampattern, resulting in poor data transmission performance. To address this problem, two novel schemes \cite{du2021integrated} have been proposed, which are also depicted in Fig. \ref{fig7}, with their tailored frame structures in Fig. \ref{fig4}.

    \begin{itemize}
    	\item \textbf{ISAC with Dynamic Beam (ISAC-DB):} 
    	An intuitionistic idea is to adjust the beamwidth dynamically to ensure that the entire vehicle geometry can be illuminated. This would result in the resolved scatters being almost uniformly distributed throughout the vehicle geometry, allowing us to infer the position of the communication antenna. However, widening the beam can lead to a lower array gain, which is not beneficial for communication transmission.
    	
    	\item \textbf{ISAC with Alternant Wide Beam and Narrow Beam (ISAC-AB):}
		To improve the communication rate, an upgraded scheme was proposed to split each transmission block into two stages. The first stage employs the dynamic beamwidth described earlier, while the second stage uses mMIMO narrow beam transmission aided by the estimated angle from the first stage. In some sense, the dynamic part of the first stage plays a similar role to pilots, but also has communication ability. We also construct an optimal duration allocation scheme as a convex optimization problem. The objective is to maximize the average achievable rate in each block. Thanks to this frame-level optimization, the resultant communication performance is much more reliable and robust compared to communication-based counterparts, especially in high-mobility environments.
  	\end{itemize}


	Note that the beamwidth can be adjusted by activating different transmit antennas, as demonstrated in \cite{du2021integrated}, through setting zero power on a transmit sub-array. However, this measure is not cost-effective since each antenna suffers a larger power burden when the transmit antennas are employed less frequently. Therefore, the transmit array requires high-power amplifiers, which are more expensive. To mitigate this adverse effect, we may rely on optimization-based beamforming design, in which case all available transmit antennas can be employed with equal transmit powers (corresponding to multiple low power amplifiers), and the beamwidth can also be adjusted according to system requirements. From another perspective, one aims to expand the beamwidth to illuminate the entire vehicle geometry. However, a narrow beam sweeping the extended vehicle target in fast time can also achieve a similar effect. This approach can be realized using the random beamforming technique, which deserves further study.
	
	\subsection{{Multi-Target Association}}
	In practical applications, the RSU should serve multiple vehicles with multiple beams simultaneously. However, one critical issue is the beam association on the multiple data streams. Let's assume that the RSU collects echo measurements from vehicles. In this case, the task of beam association is to establish the corresponding relationship between the measurement set and the vehicle identity (ID) set. This relationship directly affects whether the RSU can transmit the correct data through the predictive beamformer to the intended vehicle. If the beam association fails, then the beam tracking may fail, due to significantly larger prediction errors. To address this issue, advanced data association techniques like the well-known joint probabilistic data association (JPDA) filter and multiple hypothesis tracking (MHT), etc. may be used. For a rapid and simple implementation, two association metrics have also been employed and specifically developed in \cite{liu2020TWC, wang2022multi}.
	
    \begin{itemize}
	\item \textbf{Euclidean Distance:} Euclidean distance measures the length of a line segment between the two points in Euclidean space. When the predicted state of the vehicle is obtained via Kalman filters, then its measurement prediction can also be constructed with the measurement mapping relationship. The beam association is therefore completed by minimizing the Euclidean distance between the measurement prediction and the real counterpart \cite{liu2020TWC}. However, this scheme has ignored the uncertainties of vehicle locations, since the statistical features of measurements are not exploited.
	\item \textbf{Kullback-Leibler (KL) Divergence:} KL divergence (also known as relative entropy) is a statistical distance, which measures the difference between two probability distributions. By fully considering the estimation uncertainty, the KL divergence between state probability distributions respectively conditional to estimated and predicted states, can be calculated in Gaussian forms. Finally, the beam association is completed by minimizing KL divergence. Thanks to the exploitation of statistical uncertainties, KL divergence-based scheme has a better ID association performance than Euclidean distance-based scheme \cite{wang2022multi}.
	\end{itemize}

	\subsection{{Complex Road Geometry}}
	Current beam tracking enabled by ISAC still primarily focuses on regular road geometries, leaving broader scenarios outside the scope of effective handling. In light of this limitation, existing beam tracking approaches need to be developed specifically to accommodate arbitrarily shaped roads. Towards this end, three feasible schemes can be summarized as follows.

    \begin{itemize}
	\item \textbf{Constant Angular Velocity Model:} The beam tracking method in the angular domain has been studied for the highly non-linear mobile mmWave channel \cite{lim2019beam}. Specifically, this method assumes that the angular velocity of the target is constant, allowing for tracking without roadway information. However, in most common scenarios, the angular velocity is not constant, leading to significant approximation errors. Additionally, the method disregards the DoF of distance, resulting in a relative loss of localization information.
	\item \textbf{State-Free Model: } In this method, the kinematic function is decomposed in the Cartesian coordinate system. The position of the vehicle is then predicted coarsely based solely on the measurements attained from the past 3 data frames \cite{liu2020part3}, without taking into account the state evolution model. While this approach shows good tracking performance in high-SNR regimes and is simple to implement, it can result in unacceptable predictive errors in low-SNR regimes since the statistical properties of noise are ignored.
	\item \textbf{Curvilinear Coordinate System (CSS) Model:} To address the challenge in modeling the complex vehicle motion using Cartesian coordinate system, one may rely on the CCS model that is straightforwardly applied to irregular roadway geometry. This model represents three coordinate axes based on a curved geometry: the distance travelled along the curvature of an arbitrary road, the lateral distance, and the perpendicular distance. The interplay between the CSS system and the Cartesian coordinate system is determined by the cubic spline interpolation algorithm, which achieves trajectory fitting with high accuracy. Overall, the application of the CSS system enables superior and robust beam tracking performance, thanks to the more precise characterization and fitting equations of the road geometry \cite{meng2022vehicular}.
    \end{itemize}

\begin{figure}[!t]
	\subfigure[Simulation scenario.] { \label{fig5a}
		\includegraphics[width=1\columnwidth]{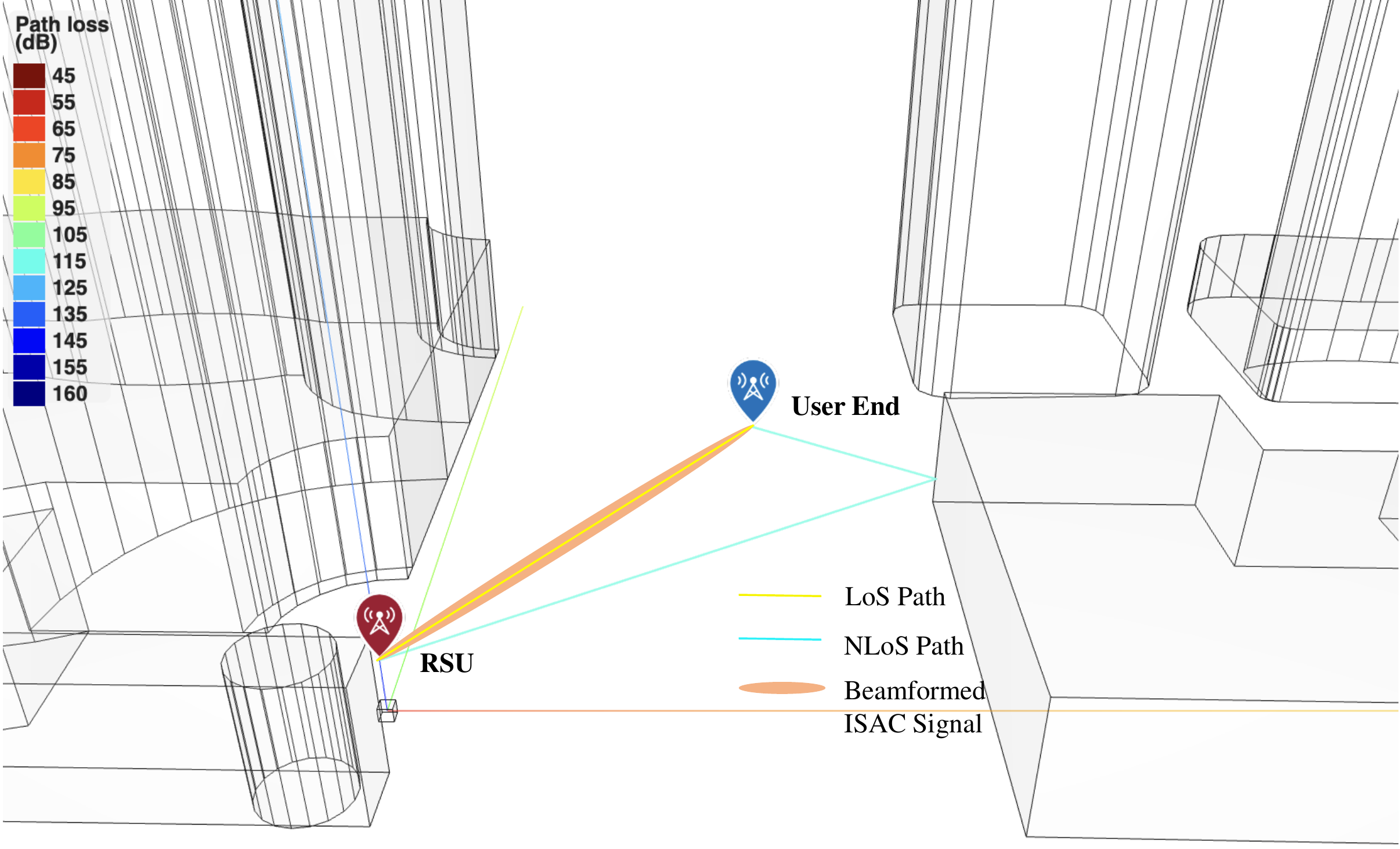}
	}
	\subfigure[Frame in simulations: conventional NR frame and ISAC NR frame.] { \label{fig5b}
		\includegraphics[width=1\columnwidth]{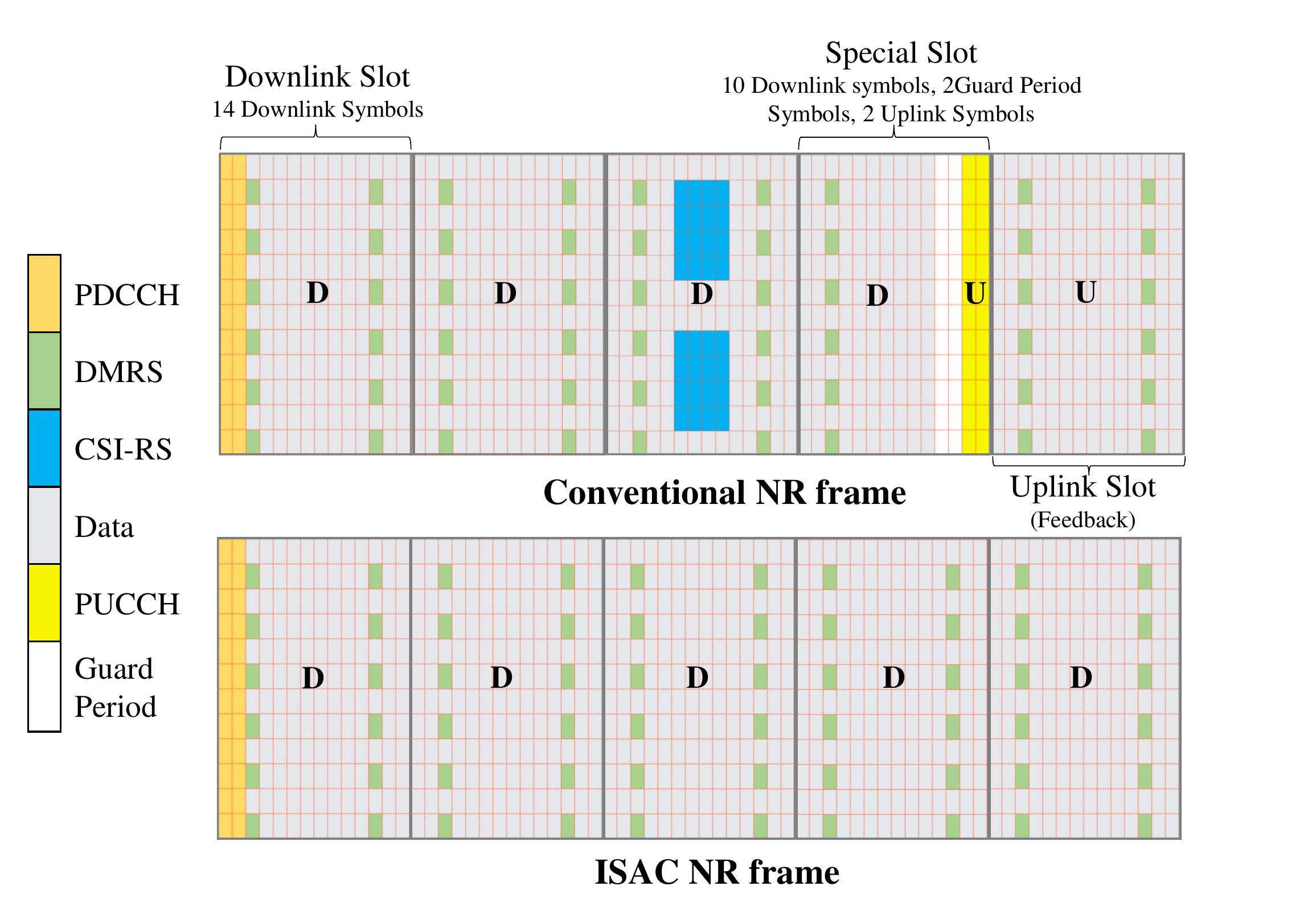}
	}
	\subfigure[Throughput comparison between ISAC and communication signals, where $N_t$ and $M_r$ denote the number of transmit ISAC antennas and receive communication antennas.] { \label{fig5c}
		\includegraphics[width=1\columnwidth]{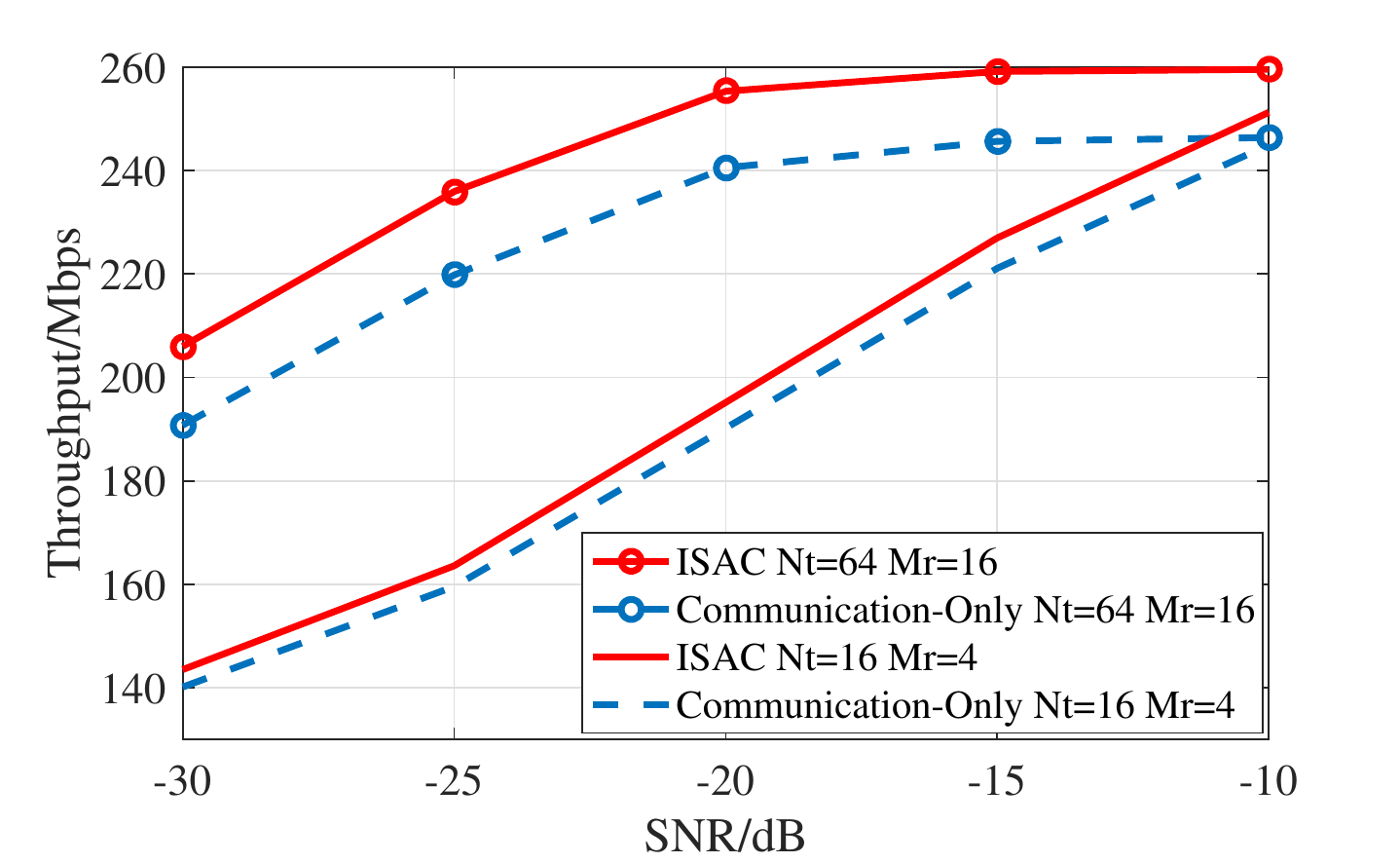}
	}
	\caption{ISAC enabled V2I transmission based on 5G NR.}
	\label{fig}
\end{figure}


\section{Case Study: 5G NR in V2I Networks}\label{sec4}
To verify the better reliability and robustness of ISAC-enabled beam tracking, a case study of utilizing 5G NR waveform in V2I networks is elaborated. We commence with the detailed beam tracking procedure of conventional NR waveform, and further concentrate on the benefits of introducing ISAC capabilities into this procedure, which has been carried out in a ray-tracing simulation for a fair comparison.

\subsection{Conventional NR Beam Tracking}
During the initial access phase in 5G NR, the gNodeB transmits a set of SSBs, with each SSB pointing towards different directions. The user end (UE) then measures the reference signal received power (RSRP) of each SSB and determines the best transmit beam, which it feeds back to the gNodeB. The gNodeB then transmits a more refined CSI-RS beam, and the UE goes through the same procedure to determine the best receive beam via beam sweeping.
Once the initial access phase is complete, the UE operates in the connected mode. 

\subsection{ISAC NR Beam Tracking}
Exploiting the ISAC signal can improve the overall throughput by releasing overheads of some specific reference signals. For example, in a single user MIMO (SU-MIMO) V2I case, the CSI-RS report contains channel information such as rank indicator (RI), precoding matrix indicator (PMI), and channel quality information (CQI). The PMI and CQI indicate the precoding direction and the channel quality, respectively, but these can be reduced with sensing-assisted communications.
Specifically, the UE's motion parameters, such as distance, velocity, azimuth angle, and elevation angle, can be estimated from the measured echoes using 2D-FFT and 2D-multiple signal classification (MUSIC) algorithms. Finally, the state prediction can be achieved via the Kalman filtering algorithm.

We utilized a ray-tracing based propagation simulator to simulate realistic road scenarios, as shown in Fig. \ref{fig5a}. The simulation incorporates NR frames and parameters, which are illustrated in Fig. \ref{fig5b}, highlighting the reduction in overheads resulting from the exploitation of the ISAC technique instead of CSI-RS and uplink feedback. Thanks to these benefits, the throughput of the ISAC-enabled signal is generally improved relative to the communication-only benchmark, as demonstrated in Fig. \ref{fig5c} \cite{li5GNR}.


\section{Open Problems and Opportunities}
Despite the proposed sensing-assisted communications technique exhibiting a promising capability of fast and robust beam tracking, there still remain several open challenges to address.
    \subsection{{High-Mobility Effect}}
	High-mobility presents a significant challenge for ISAC systems. Time-varying S\&C channels introduce several issues, including the serious ICI effect when Doppler shifts of all subcarriers differ considerably in each symbol duration. Furthermore, assuming the target is static in the coherent time interval may be unrealistic, as high-mobility results in range migration that aggravates the ICI effect. Additionally, the communication antenna is easier to go beyond the mainlobe of the beampattern, resulting in a higher probability of beam misalignment. To mitigate adverse effects, selecting OTFS as the transmit waveform or devising receive algorithms for the complete echo model involving ICI effects are underlying solutions.

	\subsection{{Driving Behavior Cognition}}
	In addition to the high-mobility of vehicles, time-varying driving behaviors such as overtaking and swerving can also cause high dynamics in S\&C channels. To address this challenge, the relationship between the state variables (such as angle and distance) and driving behaviors should be fully exploited in the beam tracking process.
	In V2I scenarios, it is particularly important to consider the varying lateral velocity of lanes, which makes the classic kinetics model with constant velocities on a straight road no longer applicable. To account for multiple driving behaviors, a multi-hypothesis model could be developed in conjunction with the interacting multiple model (IMM) \cite{meng2022vehicular}-based Kalman filtering or Bayesian filtering.

	\subsection{{NLoS Exploitation and Clutter Suppression}}\label{subsec5A} 
	As mentioned earlier, the spatial DoF referred to the total number of propagation paths in the S\&C channels, is an important performance metric in ISAC scenarios, especially for the V2I networks. In particular, multipath, including LoS/NLoS paths, can all contribute to communication transmission, but most of them are useless for sensing targets. The communication receiver on the vehicle collects NLoS components for information transmission, while the sensing receiver also receives undesired clutter from NLoS scatterers.
	Given this tradeoff relationship between S\&C channels in ISAC, an appropriate resource allocation scheme may be devised to improve the overall behaviors of beam tracking.
	
	\subsection{{Multi-Extended Target Association: Shape Estimation}}
	The side information regarding a vehicle's shape and extension range can significantly aid in target recognition. However, the predictive beam tracking method described in \cite{du2021integrated} is not capable of handling such information. This can result in considerable localization errors, especially when dealing with vehicles of irregular shapes. Moreover, tracking multiple extended targets simultaneously becomes a more challenging task. Recent research suggests that this issue can be addressed by adopting the classic framework of random finite sets (RFS). One specific implementation of this approach is the Poisson Multi-Bernoulli Mixture (PMBM) filter \cite{williams2015marginal}.

\section{Conclusion}
We present a systematic review of ISAC-empowered V2I networks. We commence with an overview of the general system architecture, which includes the platform architecture, transmit waveforms, S\&C channels, and receiving algorithms, among others. We then elaborate in detail on four challenging issues that have been studied in the literature.
Additionally, we provide a system-level case study of utilizing 5G NR waveform for predictive beam tracking simulation, which reveals the potential of ISAC in V2I networks. We also summarize open problems unsolved, highlighting research directions ahead.
Overall, this comprehensive review sheds new light on the burgeoning ISAC area, especially its significant role in the future CAV-based metaverse applications.

\ifCLASSOPTIONcaptionsoff
\newpage
\fi

\bibliographystyle{IEEEtran}
\bibliography{reference.bib}

\begin{thebibliography}{10}
\providecommand{\url}[1]{#1}
\csname url@samestyle\endcsname
\providecommand{\newblock}{\relax}
\providecommand{\bibinfo}[2]{#2}
\providecommand{\BIBentrySTDinterwordspacing}{\spaceskip=0pt\relax}
\providecommand{\BIBentryALTinterwordstretchfactor}{4}
\providecommand{\BIBentryALTinterwordspacing}{\spaceskip=\fontdimen2\font plus
\BIBentryALTinterwordstretchfactor\fontdimen3\font minus
  \fontdimen4\font\relax}
\providecommand{\BIBforeignlanguage}[2]{{%
\expandafter\ifx\csname l@#1\endcsname\relax
\typeout{** WARNING: IEEEtran.bst: No hyphenation pattern has been}%
\typeout{** loaded for the language `#1'. Using the pattern for}%
\typeout{** the default language instead.}%
\else
\language=\csname l@#1\endcsname
\fi
#2}}
\providecommand{\BIBdecl}{\relax}
\BIBdecl

\bibitem{tang2022roadmap}
F.~Tang, X.~Chen, M.~Zhao, and N.~Kato, ``The roadmap of communication and
  networking in 6{G} for the metaverse,'' \emph{IEEE Wireless Commun}, 2022.

\bibitem{cui2021integrating}
Y.~Cui, F.~Liu, X.~Jing, and J.~Mu, ``Integrating sensing and communications
  for ubiquitous iot: Applications, trends, and challenges,'' \emph{IEEE Netw},
  vol.~35, no.~5, pp. 158--167, 2021.

\bibitem{2016Millimeter}
J.~Choi, V.~Va, N.~Gonzalez-Prelcic, R.~Daniels, C.~R. Bhat, and R.~W. Heath,
  ``Millimeter-wave vehicular communication to support massive automotive
  sensing,'' \emph{IEEE Commun Mag}, vol.~54, no.~12, pp. 160--167, 2016.

\bibitem{liu2020TWC}
F.~Liu, W.~Yuan, C.~Masouros, and J.~Yuan, ``Radar-assisted predictive
  beamforming for vehicular links: Communication served by sensing,''
  \emph{IEEE Trans. Wireless Commun}, vol.~19, no.~11, pp. 7704--7719, 2020.

\bibitem{yuan2020}
W.~Yuan, F.~Liu, C.~Masouros, J.~Yuan, D.~W.~K. Ng, and
  N.~Gonz{\'a}lez-Prelcic, ``Bayesian predictive beamforming for vehicular
  networks: A low-overhead joint radar-communication approach,'' \emph{IEEE
  Trans. Wireless Commun}, vol.~20, no.~3, pp. 1442--1456, 2020.

\bibitem{yuan2021integrated}
W.~Yuan, Z.~Wei, S.~Li, J.~Yuan, and D.~W.~K. Ng, ``Integrated sensing and
  communication-assisted orthogonal time frequency space transmission for
  vehicular networks,'' \emph{IEEE J. Sel Topics. Signal Process}, vol.~15,
  no.~6, pp. 1515--1528, 2021.

\bibitem{du2021integrated}
Z.~Du, F.~Liu, W.~Yuan, C.~Masouros, Z.~Zhang, S.~Xia, and G.~Caire,
  ``Integrated sensing and communications for {V2I} networks: Dynamic
  predictive beamforming for extended vehicle targets,'' \emph{IEEE Trans.
  Wireless Commun}, 2022.

\bibitem{meng2022vehicular}
X.~Meng, F.~Liu, C.~Masouros, W.~Yuan, Q.~Zhang, and Z.~Feng, ``Vehicular
  connectivity on complex trajectories: Roadway-geometry aware {ISAC}
  beam-tracking,'' \emph{IEEE Trans. Wireless Commun}, 2022.

\bibitem{barneto2019full}
C.~B. Barneto, T.~Riihonen, M.~Turunen, L.~Anttila, M.~Fleischer, K.~Stadius,
  J.~Ryyn{\"a}nen, and M.~Valkama, ``Full-duplex ofdm radar with lte and 5g nr
  waveforms: Challenges, solutions, and measurements,'' \emph{IEEE Trans.
  Microw. Theory Techn}, vol.~67, no.~10, pp. 4042--4054, 2019.

\bibitem{dong2022sensing}
F.~Dong, F.~Liu, Y.~Cui, W.~Wang, K.~Han, and Z.~Wang, ``Sensing as a service
  in 6{G} perceptive networks: A unified framework for {ISAC} resource
  allocation,'' \emph{IEEE Trans. Wireless Commun}, 2022.

\bibitem{wang2022multi}
Z.~Wang, K.~Han, J.~Jiang, F.~Liu, and W.~Yuan, ``Multi-vehicle tracking and
  {ID} association based on integrated sensing and communication signaling,''
  \emph{IEEE Wireless Commun. Lett}, vol.~11, no.~9, pp. 1960--1964, 2022.

\bibitem{lim2019beam}
J.~Lim, H.-M. Park, and D.~Hong, ``Beam tracking under highly nonlinear mobile
  millimeter-wave channel,'' \emph{IEEE Commun. Lett}, vol.~23, no.~3, pp.
  450--453, 2019.

\bibitem{liu2020part3}
F.~Liu and C.~Masouros, ``A tutorial on joint radar and communication
  transmission for vehicular networks-part {III}: Predictive beamforming
  without state models,'' \emph{IEEE Commun. Lett}, 2020.

\bibitem{li5GNR}
Y.~Li, F.~Liu, Z.~Du, W.~Yuan, and C.~Masouros, ``{ISAC}-enabled {V2I} networks
  based on 5{G NR}: How many overheads can be reduced?'' in \emph{IEEE
  International Conference on Communications}.\hskip 1em plus 0.5em minus
  0.4em\relax IEEE, 2023.

\bibitem{williams2015marginal}
J.~L. Williams, ``Marginal multi-{B}ernoulli filters: {RFS} derivation of
  {MHT}, {JIPDA}, and association-based {M}e{MB}er,'' \emph{IEEE Trans. Aerosp.
  Electron. Syst}, vol.~51, no.~3, pp. 1664--1687, 2015.

\end{thebibliography}

\end{document}